\documentclass[sigconf]{acmart}

\usepackage{amsfonts,amssymb}

\newcommand{\ignore}[1]{}
\usepackage{xspace}
\newcommand{\paratitle}[1]{\vspace{1.5ex}\noindent\textbf{#1}}
\newcommand{\ie}{\emph{i.e.,}\xspace}

\newcommand{\eg}{\emph{e.g.,}\xspace}

\newcommand{\ourname}{SEATER\xspace}

\newcommand{\mcal}[1]{\mathcal{#1}}
\newcommand{\mrm}[1]{\mathrm{#1}}
\newcommand{\mbf}[1]{\mathbf{#1}}
\newcommand{\mbb}[1]{\mathbb{#1}}

\usepackage{amsmath}
\usepackage{multirow}
\usepackage{multicol}
\usepackage{caption}
\usepackage{subcaption}
\usepackage{latexsym}
\usepackage{mflogo}
\usepackage{graphics}
\usepackage{tabularx}
\usepackage{threeparttable}
\usepackage{balance}
\usepackage[linesnumbered,ruled,vlined]{algorithm2e}
\AtBeginDocument{%
  \providecommand\BibTeX{{%
    \normalfont B\kern-0.5em{\scshape i\kern-0.25em b}\kern-0.8em\TeX}}}

\author{Zihua Si}
\affiliation{
  \institution{Renmin University of China}
  \city{Beijing}\country{China}
  }
\email{zihua_si@ruc.edu.cn}

\author{Zhongxiang Sun}
\affiliation{
  \institution{Renmin University of China}
  \city{Beijing}\country{China}
  }
\email{sunzhongxiang@ruc.edu.cn}

\author{Jiale Chen}
\affiliation{%
  \institution{Kuaishou Technology Co., Ltd.}
  \city{Beijing}\country{China}
  }
\email{chenjiale@kuaishou.com}

\author{Guozhang Chen}
\affiliation{%
  \institution{Kuaishou Technology Co., Ltd.}
  \city{Beijing}\country{China}
  }
\email{chenguozhang@kuaishou.com}

\author{Xiaoxue Zang}
\affiliation{%
  \institution{Kuaishou Technology Co., Ltd.}
  \city{Beijing}\country{China}
  }
\email{zangxiaoxue@kuaishou.com}

\author{Kai Zheng}
\affiliation{%
  \institution{Kuaishou Technology Co., Ltd.}
  \city{Beijing}\country{China}
  }
\email{zhengkai@kuaishou.com}

\author{Yang Song}
\affiliation{%
  \institution{Kuaishou Technology Co., Ltd.}
  \city{Beijing}\country{China}
  }
\email{ys@sonyis.me}

\author{Xiao Zhang}
\affiliation{%
  \institution{Renmin University of China}
  \city{Beijing}\country{China}
  }
\email{zhangx89@ruc.edu.cn}

\author{Jun Xu}
\authornote{The corresponding author. Work partially done at Engineering Research Center of Next-Generation Intelligent Search and Recommendation, Ministry of Education.
\\
Work done when Zihua Si and Zhongxiang Sun were interns at Kuaishou.
}
\affiliation{%
  \institution{Renmin University of China}
  \city{Beijing}\country{China}
  }
\email{junxu@ruc.edu.cn}

\author{Kun Gai}
\affiliation{%
  \institution{Unaffiliated}
  \city{Beijing}\country{China}
  }
\email{gai.kun@qq.com}

\renewcommand{\shortauthors}{Zihua Si et al.}

\begin{document}




\title{Generative Retrieval with Semantic Tree-Structured Identifiers and Contrastive Learning}

\begin{abstract}
In recommender systems, the retrieval phase is at the first stage and of paramount importance, requiring both effectiveness and \textit{very high} efficiency. Recently, generative retrieval methods such as DSI and NCI, offering the benefit of end-to-end differentiability, have become an emerging paradigm for document retrieval with notable performance improvement, suggesting their potential applicability in recommendation scenarios. A fundamental limitation of these methods is their approach of generating item identifiers as text inputs, which fails to capture the intrinsic semantics of item identifiers as indices. The structural aspects of identifiers are only considered in construction and ignored during training. In addition, generative retrieval methods often generate imbalanced tree structures and yield identifiers with inconsistent lengths, leading to increased item inference time and sub-optimal performance. We introduce a novel generative retrieval framework named \ourname,  which learns \textbf{SE}m\textbf{A}ntic \textbf{T}ree-structured item identifi\textbf{ER}s using an encoder-decoder structure. 
To optimize the structure of item identifiers, \ourname incorporates two contrastive learning tasks to ensure the alignment of token embeddings and the ranking orders of similar identifiers. 
In addition, \ourname devises a balanced $k$-ary tree structure of item identifiers, thus ensuring consistent semantic granularity and inference efficiency. 
Extensive experiments on three public datasets and an industrial dataset have demonstrated that \ourname outperforms a number of state-of-the-art models significantly.
\end{abstract}

\ccsdesc[500]{Information systems~Recommender systems}


\keywords{Recommendation;  Generative Retrieval; Contrastive Learning}

\maketitle
\section{Introduction}
\label{sec:intro}


Modern recommendation systems (RS) predominantly use a two-stage retrieve-then-rank strategy~\cite{youtubeDNN}. 
During retrieval, a small subset of items (hundreds) is chosen from a vast item pool (millions).
Considering the large scale of the entire pool, efficiency is vital for the retrieval model. 
Moreover, the success of the ranking model depends on the quality of retrieved items, highlighting the importance of retrieval effectiveness.
Traditional models leverage dual-encoder architectures and Approximate Nearest Neighbor (ANN) algorithms.
Initially, retrieval models represent users with a single vector~\cite{youtubeDNN, SASREC}.
Subsequent studies~\cite{comirec,Re4,MIND_model} notice the inadequacy of single, finite-length vector representations, leading to the introduction of multi-vector retrieval. 
These approaches leverage multiple vectors to better express user interests and continue using ANN across multiple vectors for inference.
However, the inner product of ANN theoretically requires a strong assumption for the Euclidean space, which may not be satisfied in practical applications. 
Hence, developing a model capable of capturing complex interactions, adequately representing user interests, and ensuring efficiency is a direction worth exploring.

To achieve this goal, tree-based indexing models like TDM~\cite{TDM} and JTM~\cite{JTM19NIPS} from Alibaba have been introduced.
They estimate interaction probabilities using deep models and achieve affordable efficiency by retrieving over tree-based indices, rather than the entire item pool.
A recent trend in search~\cite{DSI, NCI} views retrieval as a generation task. 
These models use transformer memory as a differentiable index and decode document IDs autoregressively as texts.
The latest work TIGER~\cite{rajput2023Tiger} utilizes similar architectures for RS. They construct codebooks (so-called identifiers in this paper) to represent items and estimate interaction probabilities using the product probabilities predicted by transformers.


Despite their achievements, these generative models cannot meet the efficiency requirement for large-scale applications and their performance can be improved.
Regarding effectiveness, these works treat identifier tokens purely as texts, optimizing with cross-entropy loss loosely related to the indexing structures, neglecting the inherent characteristics of such structures.
The structural aspects are only considered when constructing identifiers and are not integrated into the loss function during training.
In terms of efficiency, imbalanced tree-structured identifiers in DSI and NCI can result in increased and inconsistent inference time for items. The multiple transformer layers in TIGER increase the computational burden during inference. 

To address such problems, we propose a generative model for the recommendation, namely \ourname, which learns \textbf{SE}m\textbf{A}ntic \textbf{T}ree-structured item identifi\textbf{ER}s via contrastive learning.
We leverage an encoder-decoder model which encodes user interests and decodes probably the next items. 
The decoder represents items into equal-length identifiers with consistent semantics within the same level. 
\ourname assigns balanced $k$-ary tree-structured identifiers to items and learns semantics and hierarchies of identifier tokens through contrastive learning tasks.
We construct such identifiers based on collaborative filtering information to incorporate prior knowledge. 
During training, we design two contrastive learning tasks to help the model comprehend the structure of item identifiers.
Considering that each token represents an individual set of items, each identifier token has distinct semantics. 
The hierarchical relationship and inter-token dependency are inherent properties of such tree-structured indices. 
However, relying solely on user-item interactions for learning these complicated associations is challenging.
It is necessary to introduce additional tasks to learn this structural information.
We integrate two contrastive learning tasks in addition to the generation task.
The first task employs the infoNCE loss, aligning token embeddings based on their hierarchical positions. 
The second task leverages a triplet loss, instructing the model to differentiate between similar identifiers.
In this way, \ourname obtains both efficiency and effectiveness for item retrieval in RS.
Extensive experiments across four datasets validate the effectiveness of the proposed model.

In summary, our main contributions are as follows:

\noindent$\bullet$ We introduce a generative framework, \ourname, for the retrieval phase of recommendation.
We elaborate on the construction of identifiers, structural optimization based on contrastive learning.


\noindent$\bullet$ Utilizing two contrastive learning tasks, the model captures the semantics of the tokens and the hierarchies within the tree structure. 
Both tasks optimize identifiers' structures.

\noindent$\bullet$ The balanced $k$-ary tree structure ensures consistent semantic granularity for tokens at the same level and significantly reduces inference time compared with other tree-structured methods.

\noindent$\bullet$ Extensive experiments\footnote{Implementations available at this link (\textcolor{magenta}{\url{https://github.com/Ethan00Si/SEATER_Generative_Retrieval}}).} on three public datasets and an industrial dataset have demonstrated that \ourname significantly outperforms several state-of-the-art (SOTA) methods, including dual-encoder, tree-based indexing, and generative methods. 
\section{RELATED WORK}

\paratitle{Retrieval in Recommender Systems.}
In RS, the retrieval phase selects a subset of items from a vast corpus. 
For efficiency, the industry often uses dual-encoder models to represent users and items as vectors~\cite{youtubeDNN, MIND_model, comirec, UMI2022, SDM19, Re4}.
And user preferences towards items are estimated through the inner product of vectors, which can be accelerated by ANN search for inference.
The initial dual-encoder models represented users with a single vector~\cite{youtubeDNN, SASREC, GRU4REC}. Subsequent studies~\cite{MIND_model, comirec, UMI2022, Re4},  observed the limitations of expressing with a finite-length single vector and introduced multi-vector user interest modeling, continuing to utilize ANN search for inference.
An alternate research direction aims to enable more intricate models with complex interaction estimation. 
TDM~\cite{TDM} and JTM~\cite{JTM19NIPS} proposed by Alibaba involve tree-based indexing with advanced deep models, thereby facilitating more accurate estimation.
RecForest~\cite{NIPS22RecForest} constructs a forest by creating multiple trees and integrates a transformer-based structure for routing operations.
Similar to those studies, this paper seeks to retrieve items in a generative manner and optimize item identifiers from the indices perspective.

\paratitle{Generative Retrieval.}
In document retrieval, researchers have investigated using pre-trained language models to generate various types of document identifiers.
For example, DSI~\cite{DSI} and NCI~\cite{NCI} utilize the T5~\cite{2020t5} model to produce hierarchical document IDs, while SEAL~\cite{SEAL2022autoregressive} (with BART~\cite{lewis-etal-2020-bart} backbone) and ULTRON~\cite{zhou2022ultron} (using T5) use titles or substrings as identifiers. AutoTSG~\cite{zhang2023termsets} and NOVO~\cite{TermSet_Gen_Rtrieval_cikm23}, also based on T5, employ term-sets and n-gram sets as identifiers.
There are studies exploring the identifier structures, such as GenRet~\cite{NEURIPS2023_91228b94} and LTRGR~\cite{li2023learning}.
Generative document retrieval has been expanded to various fields. IRGen~\cite{zhang2023irgen} uses a ViT-based model for image search, while TIGER~\cite{rajput2023Tiger} employs the T5-based architecture for RS.
However, due to the \emph{resource-intensive} nature of multiple transformer layers, these studies are ill-suited for large-scale item retrieval in RS. 
Different from them, this paper delves into the use of more \emph{parameter-efficient} models for generative retrieval in such systems.
Also, there are previous works utilizing the generative nature of language models for recommendation, including P5~\cite{P5_zhangyongfeng, hua2023index},  TIGER~\cite{rajput2023Tiger}, and GPTRec~\cite{petrov2023GPTRec}. These approaches, after establishing item identifiers, also known as codebooks, do not optimize these identifiers' structures. Ideally, the models would optimize the identifiers related to corresponding indices. Towards this end, this paper learns the inherent hierarchy and relationships of identifier structures with the help of contrastive learning.

\section{Method}
In this section, we elaborate on the proposed model, detailing its model architecture, training, and inference stages.

\subsection{Overview}
Suppose a user $u\in \mathcal{U}$ accesses a retrieval system, and the system returns a list of candidates with each item $v\in\mathcal{I}$, where $\mathcal{U}$ and $\mathcal{I}$ denote the entire sets of users and items respectively. 
Let's use $\boldsymbol{x} =[x_1, \cdots, x_t]\in\mathcal{X}$ to denote the historically interacted items of user $u$. In generative retrieval, the identifier of each item $v\in \mathcal{I}$ is represented as a token sequence $\boldsymbol{y}=[y_1, \cdots, y_l]\in\mathcal{Y}$, where $l$ is the length of the identifier.  
The goal of the generative retrieval model is learning a mapping $f:\mcal{X}\rightarrow \mcal{Y}$, which takes a user's interacted item sequence as input and generates a sequence of tokens (candidate identifiers).

\begin{figure}[t]
    \centering
    \includegraphics[width=0.95 \linewidth]{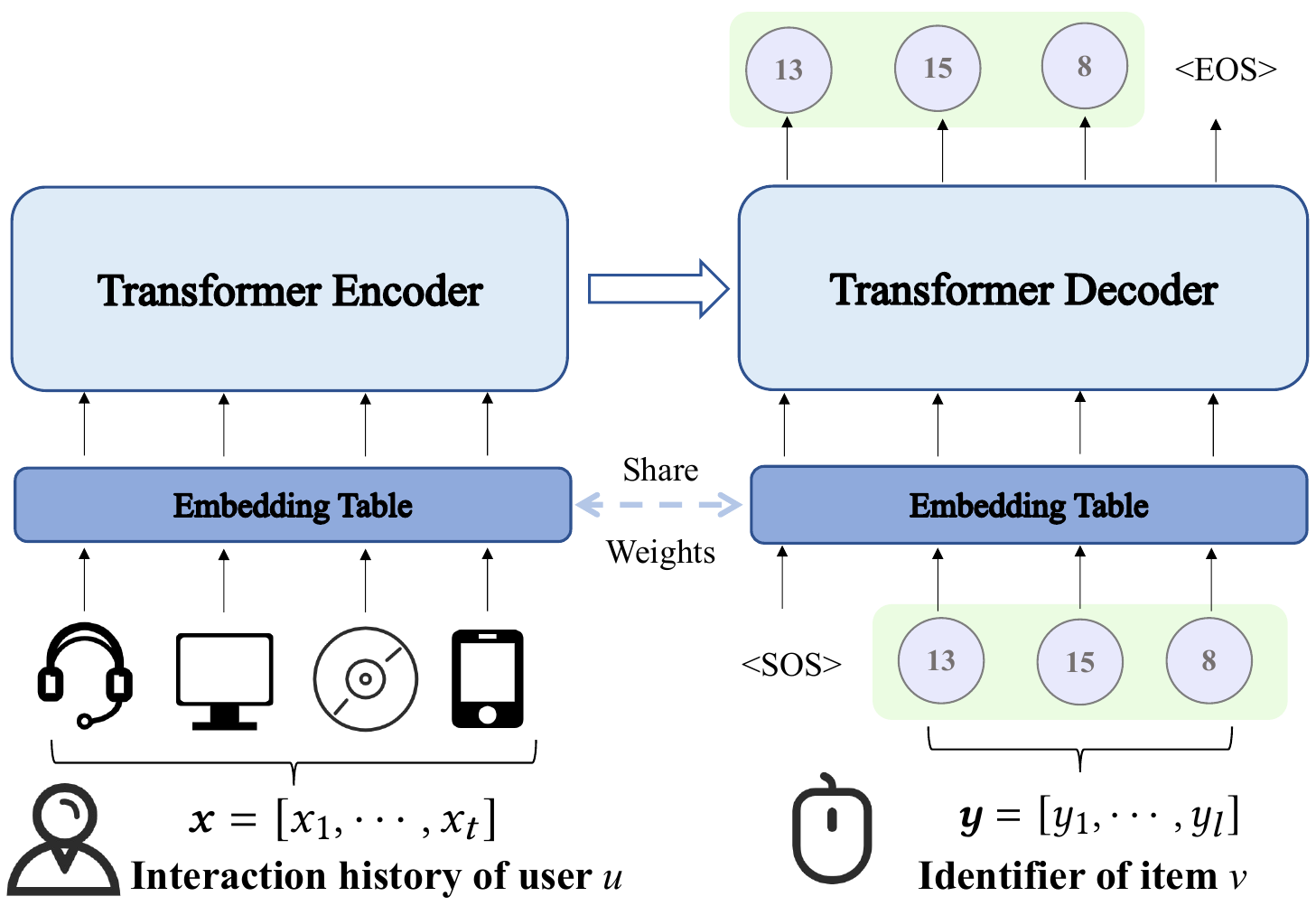}
    \vspace{-5px}
    \caption{A brief illustration of \ourname.
    The retrieval model encodes the interacted items $\boldsymbol{x}=[x_1,x_2,\cdots,x_t]$ of user $u$ and decodes the identifier $\boldsymbol{y}=[y_1,y_2,\cdots,y_l]$ of item $v$.
    }
    \label{fig: model}
    \vspace{-10px}
\end{figure}

As shown in~\autoref{fig: model}, the retrieval model feeds the user's behavior $\boldsymbol{x}$ into the encoder. 
Following this, the decoder employs an auto-regressive method to generate the item identifier $\boldsymbol{y}$ step by step.
The probability of interaction between user $u$ and item $v$ is estimated as:
\begin{equation}
   p(u,v) = \prod_{i=1}^{l} p(y_i|\boldsymbol{x},y_1,y_2,\dots,y_{i-1})
\end{equation}
where $l$ denotes the length of item identifiers.
To assign items with semantic representations, we convert all items into uniform-length identifiers, as depicted in~\autoref{fig: training} (a).
Identifier tokens capture item information from coarse to fine granularity, spanning from the beginning to the end.
We use a multi-task learning approach to optimize both the model and identifiers, as depicted in~\autoref{fig: training}. The sequence-to-sequence task directs the model to generate valid identifiers,
while the two contrastive learning tasks aid in grasping semantics and relationships among identifier tokens.


\subsection{Retrieval Model}

\subsubsection{Encoder-Decoder Architecture}

For the retrieval model, we employ the standard Transformer architecture~\cite{DBLP:journals/corr/VaswaniSPUJGKP17}. Detailed Transformer structure specifics are omitted for brevity.

We leverage the Transformer encoder to capture user interests from behavior sequences: 
\begin{equation}
    \mbf{X} = \mrm{Encoder}(x_1,x_2,\cdots,x_t),
    \label{eq: encoder}
\end{equation}
where $\mbf{X} \in \mbb{R}^{t\times d}$ denotes the encoder hidden states of user interaction history $\boldsymbol{x}=[x_1,x_2,\cdots,x_t]$, $t$ denotes the number of interacted items.
The embeddings of $t$ items serve as inputs to the encoder.

We exploit the Transformer decoder to model user-item interaction and predict the interaction probability in an auto-regressive manner. The decoder's hidden states are calculated as follows:
\begin{equation}
    \mbf{Y} = \mrm{Decoder}(\boldsymbol{x}, y_1,y_2,\dots,y_l),
    \label{eq: decoder}
\end{equation}
where $\mbf{Y} \in \mbb{R}^{l\times d}$ denotes the decoder hidden states of item identifier $\boldsymbol{y}=[y_1,y_2,\cdots,y_l]$, $l$ is the length of the item identifier.
The embeddings of identifiers serve as inputs to the decoder.

In our study,  we refrained from stacking numerous Transformer layers, \eg 12 Transformer blocks in T5-Base~\cite{2020t5}.
We used \textbf{just one layer} to maintain efficiency in large-scale item retrieval contexts. In Section~\ref{Appendix: parameter count}, we show that more layers indeed help the performance, with potential loss of efficiency.

The probability at step $i$ can be modeled by softmax value of the $i$-th decoder hidden state $\mbf{y}_i$ and candidate tokens $\mcal{C}$:
\begin{equation}
    p(y_i|\boldsymbol{x},y_1,y_2,\dots,y_{i-1}) = \frac{ \exp( \mbf{y}_i^{\intercal} \mbf{e}_{y_i} ) }
                                        {\sum_{y_{i'} \in \mcal{C}} \exp(\mbf{y}_i^{\intercal} \mbf{e}_{y_{i'}})},
\end{equation}
where $\mbf{y}_i \in \mbf{R}^{d}$ denotes the $i$-th vector in $\mbf{Y} \in \mbb{R}^{l\times d}$, $\mbf{e}_{y_i} \in \mbf{R}^{d}$ denotes the embedding for token $y_i$, and $C$ is the set of all possible next tokens of size $k$ given the prefix $[y_1,y_2,\ldots,y_{i-1}]$.
This approach estimates interaction probability through product probabilities. The cross-attention mechanism and the decoder structure provide a comprehensive capture of interaction estimation beyond the inner product of dual-encoder models.
Likewise, user interests are represented using a matrice $\mbf X$ that incorporates the full historical sequence, as opposed to limited-length vectors. This method significantly improves the expressive power for user interests compared to the dual-encoder models.

\subsubsection{Item  Identifiers}
\label{sec: identifiers}
Considering \ourname retrieves items using identifiers, the identifiers' construction is crucial.
We have established a balanced tree structure to provide equal-length identifiers for the retrieval task, which offers numerous advantages.

\begin{algorithm}[t]
\footnotesize
\caption{Constructing equal-length identifiers.}
\label{Algo: kmeans}
\KwIn{
Item embeddings $X_{1:N}$, number of items $N$, 
number of branches $k$.
}
\KwOut{Semantic item indexes $L_{1:N}$}
\SetKwFunction{}
\KwFunction{
\textbf{Function} ConstructIdentifiers($X$) :
}
\SetAlgoLined
\BlankLine

\hspace{2em}\textcolor[rgb]{0.2,0.4,0.6}{\# Min(Max) size of each cluster}\\
\hspace{2em}MinSize $\gets \lfloor |X| / k \rfloor$, MaxSize$\gets \lfloor |X| / k \rfloor + 1$\\  
\hspace{2em}$C_{1:k} \gets$  Constrained-Kmeans($X$, MaxSize, MinSize) \\
\label{constrained k-means in algo}
\hspace{2em}$J \gets$ empty list\\

\hspace{2em}\textcolor[rgb]{0.2,0.4,0.6}{\# Recursively clustering for each cluster}\\
\hspace{2em}\textbf{For} $i=0$ \textbf{to} $k-1$ \textbf{do}\\
\hspace{3em}$J_{current} \gets$ [$i$] * |$C_{i+1}$|\\
\hspace{3em}\textbf{if} $|C_{i+1}| > k $ \textbf{then}\\
\hspace{4em}$J_{rest} \gets $ \text{ConstructIdentifiers(}$C_{i+1}$\text{)}\\
\hspace{3em}\textbf{else}\\
\hspace{4em}$J_{rest} \gets [0, \dots, |C_{i+1}| - 1]$\\
\hspace{3em}\textbf{end if}\\
\hspace{3em}$J_{cluster} \gets $ \text{ConcatString(}$J_{current}$\text{, }$ J_{rest}$\text{)}\\
\hspace{3em}$J \gets J.\text{AppendElements}(J_{cluster})$\\
\hspace{2em}$J \gets \text{ReorderToOriginal}(J,\; X,\; C_{1:k})$\\
\hspace{2em}\textcolor[rgb]{0.2,0.4,0.6}{\# Upon finishing clustering, assign unique IDs for each tree node}\\
\hspace{2em}\textbf{if} $|X| = N$ \textbf{then}\\
\label{ressign node ID in algo: start}
\hspace{3em}$i \gets N+1$, visited $\gets$ empty dict, $L \gets J$.copy()\\
\hspace{3em}\textbf{For} $r=1$ \textbf{to} $N$ \textbf{do}\\
\hspace{4em}\textcolor[rgb]{0.2,0.4,0.6}{\# Each leaf node is encoded using its item ID}\\
\hspace{4em}$L_{r,\text{last column}} \gets r$\\
\hspace{4em}\textcolor[rgb]{0.2,0.4,0.6}{\# Assign IDs to all non-leaf nodes}\\
\hspace{4em}\textbf{For} $l=1$ \textbf{to} penultimate column \textbf{do}\\
\hspace{5em}\textbf{if} $J_{r,1:l}$ \textbf{in} visited \textbf{then}\\
\hspace{6em}$L_{r,\ l} \gets$ visited( $J_{r,1:l}$ )\\
\hspace{5em}\textbf{else}\\
\hspace{6em}$L_{r,\ l} \gets i$, visited( $J_{r,1:l}$ ) $\gets i$\\
\hspace{6em}$i \gets i+1$\\
\hspace{2em}\textbf{end if}\\
\label{ressign node ID in algo: end}

\hspace{2em}\textbf{return} $L$
\end{algorithm}

\ourname utilizes a balanced $k$-ary tree structure to construct identifiers for items within set $\mcal{I}$.
To incorporate prior knowledge, we leverage a hierarchical clustering method with the constrained k-means~\cite{bennett2000constrained} algorithm, to convert items into identifiers.
Given an item set $I$ to be indexed, we recursively cluster items into equal-size $k$ groups until each group has fewer than $k$ items. 
Detailed identifier tree construction can be found in Algorithm~\ref{Algo: kmeans}. 
We employ item embeddings $X_{1:N}$ extracted from trained SASREC~\cite{SASREC} as the foundation for hierarchical clustering, leading to identifiers with collaborative filtering insights.
We assign unique tokens for each clustered node because each node represents distinct item sets.
We present a toy example to clarify our method.
As shown in~\autoref{fig: training} (a), a mouse, the $8$-th item in set $\mcal{I}$, is mapping into the identifier [\textit{13, 15, 8}], where special tokens (start and end) are omitted.
For instance, as shown in~\autoref{fig: training} (a), token `$8$' representing item $8$ and token `$15$' denoting items $7$ and $8$. 
Tokens' semantic granularity differs by layer: the leaf layer conveys item-specific details, the penultimate layer captures information from a set of $k$ items, and the topmost token embodies the whole item set's semantics.
Thus, we allocate unique embeddings to individual tokens.

Formally, we embed all identifiers and items in an embedding table $\mbf{E}\in \mbb{R}^{M\times d}$, where $M$ is the number of identifier tokens.
In other words, \emph{each tree node has unique token embedding.}
Note that the identifier tokens at the leaf layer have a one-to-one correspondence with items. 
\emph{We share embeddings between these tokens and items}.
Given $N$ as the item count, our identifiers add $(M-N)d$ extra embeddings compared to item embeddings. The fact $M-N \ll N$ causes an affordable increase in space overhead.
See Section~\ref{sec: complexity} for details.

Our construction method offers several distinct advantages:
\textbf{(1)} 
All items are mapped into equal-length identifiers due to the balanced tree structure.
The equal-length identifiers ensure that tokens at the same level possess consistent hierarchical semantics.
In an imbalanced tree, an item's identifier might end at the third level while another extends to the fifth. This causes varied semantic granularity among third-level tokens. 
Furthermore, a balanced tree ensures shorter maximum identifier lengths (tree depth) than an imbalanced tree.
This results in faster inference speed and equivalent processing time for all items, validated in Section~\ref{sec: complexity}.
\textbf{(2)}  
We build the identifier tree with item embeddings from a different retrieval model. 
Using item embeddings informed by collaborative filtering, the identifiers effectively capture prior knowledge for recommendations.
As shown in~\autoref{fig: training} (a), items like `mouse' and `computer', which have similar user interactions, tend to have similar identifiers, \ie the prefix [\textit{13, 15}].
Detailed empirical analyses in Section~\ref{sec: exp item identifiers} have verified the strengths of our semantic identifiers.
Importantly, using these embeddings doesn't add extra training loads. This aligns with industry practices where multiple models are used for multi-path retrieval, so there's no added training cost beyond \ourname.
For instance, a company may simultaneously employ models like SASRec~\cite{SASREC} and SEATER.
Hence, using SASRec embedding does not necessitate additional overhead.

\subsection{Training}
\label{sec: training}

\begin{figure}[t]
    \centering
    \includegraphics[width=0.99 \linewidth]{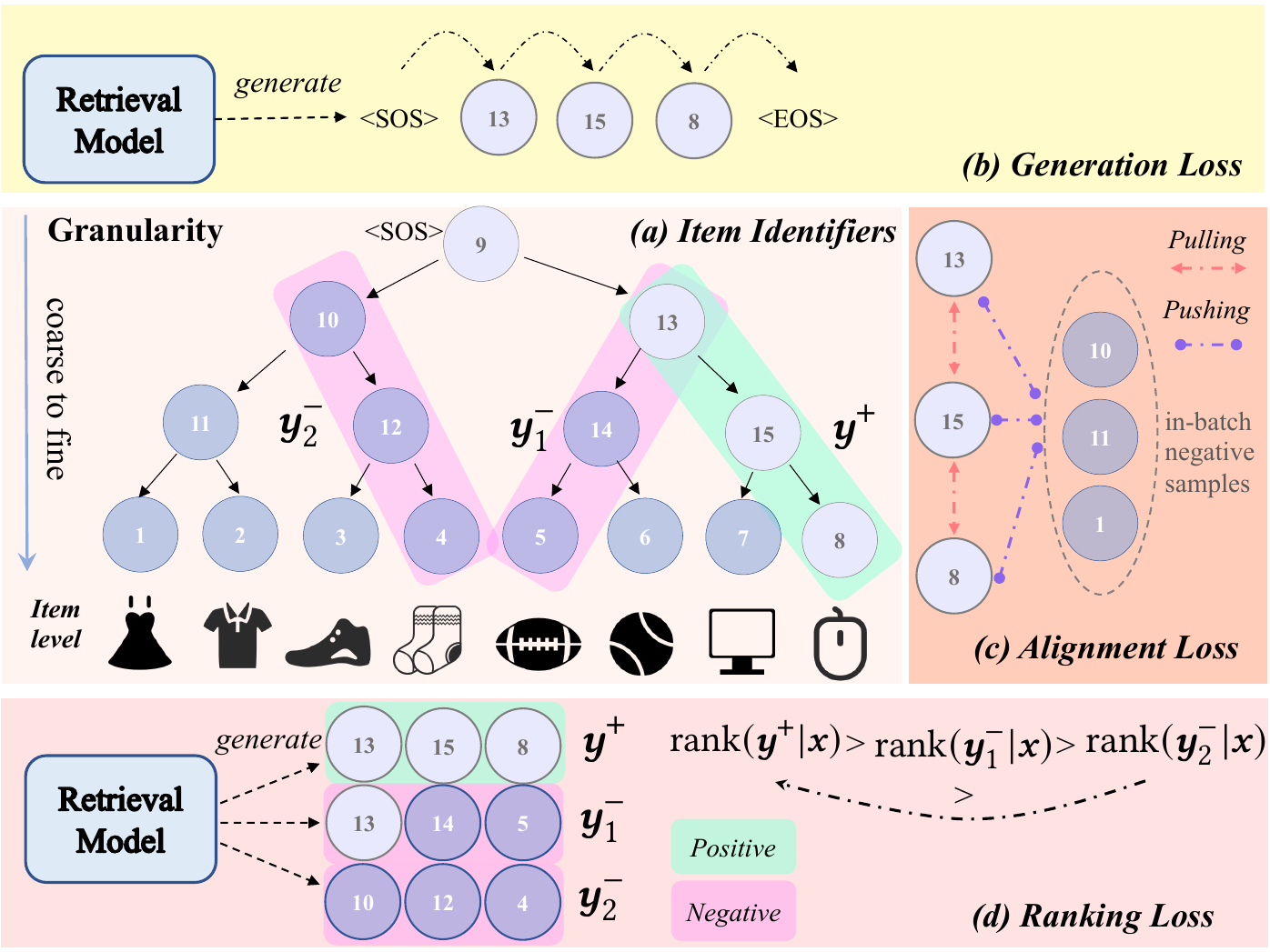 }
    \vspace{-5px}
    \caption{The proposed tree-structured identifiers and multi-task learning scheme.
    (a) An example of a balanced $k$-ary tree structure of item identifiers. Here $k$ equals 2 for simplicity. In practice, $k$ can be any integer $\geq 2$.
    `9' denotes the start token. 
    Each tree node corresponds to an unique token.
    (b)$-$(d) denote three losses for different tasks.
    (b) Generation Loss: guide the model to decode item identifiers. 
    (c) Alignment Loss: grasp semantics and hierarchies of tokens.
    (d) Ranking Loss: differentiate between similar identifiers.
    }
    \label{fig: training}
    \vspace{-15px}
\end{figure}

Traditional recommendation models rely on user-item interaction data for training. 
In generative retrieval, where both the model and indices (item identifiers) are trained, we should also consider the indices' structure for training.
To tackle this, we propose a multi-task learning scheme with two contrastive learning tasks and a generation task, as shown in~\autoref{fig: training}.

\subsubsection{Generation Loss}

We formulate the retrieval task as a sequence-to-sequence generation task for decoding item identifiers.
To generate valid item identifiers, following~\cite{NCI,DSI}, we employ the sequence-to-sequence cross-entropy loss with teacher forcing.  
As depicted in~\autoref{fig: training} (b), given a training sample $(\boldsymbol{x},\boldsymbol{y})$, the loss function can be written as follows:
\begin{equation}
    \mcal{L}_{\mrm{gen}} = - \sum_{i=1 }^{l}  \log p(y_i|\boldsymbol{x},y_1,y_2,\dots,y_{i-1}),
    \label{eq: generation loss}
\end{equation}
 where $\boldsymbol{x}$ denotes the user history, and $\boldsymbol{y}$ denotes the next interacted item's identifier.

\subsubsection{Alignment Loss}

Given that \emph{each token has distinct semantics} and \emph{inter-token relationships exist}, we utilize contrastive learning to learn identifiers from the indices perspective.

As depicted by the tree-structured identifiers in~\autoref{fig: training}, the parent token (\eg 15) encapsulates its child tokens (\eg 7 and 8), as one can only access the child tokens through the parent token.
In tree-building, we group items into $k$ clusters to form $k$ child tokens.
Thus, semantically, the parent token should align closely with the centroid of its child tokens.
For instance, token 15 represents items 7 and 8, while token 14 represents items 5 and 6.
Towards this end, we devise a contrastive learning objective shown in the~\autoref{fig: training} (c). 
 Given a token $j$, we employ the infoNCE loss to minimize the distance between it and its parent token $p$, while maximizing the distance between it and the in-batch negative instances:

\begin{equation}
    \mcal{L}_{\mrm{ali}} = - \log \frac{\exp(\cos(\mbf{e}_j,\mbf{e}_p) / \tau)}
    {\sum_{k\in \mcal{B} \setminus{j}} \exp(\cos(\mbf{e}_j, \mbf{e}_k)/ \tau)},
\end{equation}
where temperature $\tau$ is a hyper-parameter.
The parent token $p$ is viewed as the positive instance for token $j$.
Other tokens in the same batch $\mcal{B}$, excluding the child and parent of $j$, are viewed as negatives.
This loss pulls the representations of tokens with parent-child relationships closer and pushes the representations of unrelated tokens apart.

\subsubsection{Ranking Loss}
The generative model compares candidate identifiers during inference (Section~\ref{sec: inference}). 
Different identifier tokens index various items.
In this way, the model needs to \emph{discern subtle differences between similar identifiers}. 
We select identifiers with varying prefix lengths compared to the ground truth to guide the model in ranking them using a contrastive learning task.
The varying prefix lengths imply the distinction of different identifiers within hierarchies.

For each ground truth identifier $\boldsymbol{y}^+$, we randomly sample $q$ similar identifiers, denoted as $\boldsymbol{y}^-_1, \boldsymbol{y}^-_2, \cdots, \boldsymbol{y}^-_q$.
For simplicity, in $\boldsymbol{y}^-_1$, the `1' denotes the first negative sample and not the first position of the identifier.
We select $q$ samples with $q$ different shared prefix lengths from $\boldsymbol{y}^+$.
These samples indicate related items with diverse similarity levels. 
For instance, as illustrated in~\autoref{fig: training} (d) with $q=2$, $\boldsymbol{y}^-_1$ shares one token with $\boldsymbol{y}^+$, whereas $\boldsymbol{y}^-_2$ shares none.
In practice, we sample identifiers with more than two different tokens with $\boldsymbol{y}^+$.
Then, we teach the model to rank these $q+1$ identifiers.

In specific, for each sample $(\boldsymbol{x},\boldsymbol{y})$, we can get the representation vector of encoder hidden states $\mbf{z}_x \in \mbb{R}^{d}$ and decoder hidden states $\mbf{z}_y \in \mbb{R}^{d}$:
\begin{equation}
    \mbf{z}_x = \mrm{MEAN}(\mbf{X}),\quad
    \mbf{z}_y = \mrm{MEAN}(\mbf{Y}),
\end{equation}
where MEAN denotes the mean pooling, $\mbf{X}$ and $\mbf{Y}$ are obtained from equation~\ref{eq: encoder} and~\ref{eq: decoder} respectively.
After obtaining hidden states, we rank different identifiers in pairs to instruct the model on the ranking order among these $q+1$ identifiers.
The paired identifiers constitute the set $\mcal{Q}$, where $|\mcal{Q}|=C_{q+1}^2$.
For any pair $(\boldsymbol{y}^\dag, \boldsymbol{y}^\ddag)$ in $\mcal{Q}$, we rank the sample with more identical prefix tokens with $\boldsymbol{y}^+$ higher.
For example, as shown in~\autoref{fig: training} (c), we rank $\boldsymbol{y}^+$ higher than $\boldsymbol{y}^-_1$, $\boldsymbol{y}^+$ higher than $\boldsymbol{y}^-_2$, and $\boldsymbol{y}^-_1$ higher than $\boldsymbol{y}^-_2$.
In detail, we employ the triplet loss to steer the model toward learning the desired ranking orders:

\begin{equation}
    \mcal{L}_{\mrm{rank}} = \sum_{({\boldsymbol{y}^\dag, \boldsymbol{y}^\ddag)\in \mcal{Q}}}\max \left\{0, s(\mbf{z}_x, \mbf{z}_{y^\dag}) - s(\mbf{z}_x, \mbf{z}_{y^\ddag}) + \xi \right\},
    \label{eq: ranking loss}
\end{equation}
where $s$ is a similarity function, and $\xi$ denotes a positive margin value.
Here, we use $\boldsymbol{y}^\ddag$ to denote the sample with more tokens in common with $\boldsymbol{y}^+$, and $\boldsymbol{y}^\dag$ for the one with fewer.
We set $\xi$ as an adaptive value, $\xi=\beta * (\mrm{num}(\boldsymbol{y}^\ddag) - \mrm{num}(\boldsymbol{y}^\dag))$, to reflect rank differences in different pairs, where $\beta$ is a hyper-parameter set to a small positive value, and num$(\boldsymbol{y})$ denotes the number of identical tokens between $\boldsymbol{y}$ and $\boldsymbol{y}^+$.
The function $s$ is defined as: $s(\mbf{p}, \mbf{q})=\sigma(\mbf{p}^{T}\mbf{W}_s\mbf{q})$, where $\sigma$ denotes the sigmoid activation function, and the introduction of parameters $\mbf{W}_s$ ensures the similarity estimation can be more flexible.

\subsubsection{Multi-task Training}

Finally, we train our model in an end-to-end manner under a multi-task learning scheme:

\begin{equation}
    \mcal{L} = \mcal{L}_{\mrm{gen}} + \lambda_{a} \mcal{L}_{\mrm{ali}}  + \lambda_{r} \mcal{L}_{\mrm{rank}}.
\end{equation}
where $\lambda_{a}$ and $\lambda_{r}$ are hyper-parameters to balance different tasks.
We also introduce $L_2$ regularization to avoid over-fitting, which is omitted here for conciseness.


\subsection{Inference}
\label{sec: inference}
In the inference phase, our objective is to extract the top $n$ items from the entire candidate set. To achieve this, we employ a \emph{constrained beam search} mechanism on the decoder module, specifically targeting tree-based identifiers, following NCI~\cite{NCI}. 
This ensures that the model's decoding aligns with the designated prefix tree, yielding valid identifiers.


\begin{table}[t]
	\small
	\centering
	\caption{Time and space complexities analyses. 
        We consider the beam search procedure for time complexity and the size of item identifiers for space complexity.
        \ourname demonstrates superior inference speed and affordable additional space cost.
        }
        \vspace{-5px}
	\label{tab:complexity}
	\setlength{\tabcolsep}{1.9mm}{
		\begin{tabular}{l|l|l}
			\toprule
			  \textbf{Models} & \textbf{Inference Time} & \textbf{Identifier Size} \\
			\hline
			TDM~\cite{TDM} & $\mathcal{O}(b\log_2 N)$ & $\mathcal{O}(Nd)$ \\
			RecForest~\cite{NIPS22RecForest} & $\mathcal{O}(Tbk\log_k N)$ & $\mathcal{O}(Tkd)$ \\
			  DSI~\cite{DSI}  & $\mathcal{O}(bkL)$ & $\mathcal{O}(kd)$ \\
			  NCI~\cite{NCI}  & $\mathcal{O}(bkL)$ & $\mathcal{O}(kLd)$ \\
			  \ourname  &$\mathcal{O}(bk\log_k N)$ & $\mathcal{O}(Nd)$ \\
			\bottomrule
		\end{tabular}
	}
 \begin{tablenotes}
    \item[1] DSI\&NCI: In the worst case, $L$ equals $\frac{N}{k}$, resulting in $\mcal{O}(bkL)$ deteriorating to $\mcal{O}(bN)$. Empirically, $L$ is $c\log_k N$, where $c$ lies between $2$ and $4$, resulting in an inference time $c$ times that of \ourname. 
    \item[2] $N$: the number of items; $k$: the number of branches; $d$: the item embedding size; $b$: the beam size, $L$: the depth of tree in DSI and NCI; $T$: the number of trees in RecForest 
\end{tablenotes}
\vspace{-5px}
\end{table}

\section{Discussion}

In this section, we compare \ourname with related previous work.

\subsection{Comparison with Existing Work}

Generative retrieval is an emerging research direction.
We are at the forefront of incorporating the optimization of identifier structural information into the training phase of RS.

DSI~\cite{DSI} and NCI~\cite{NCI} pioneer in learning a generative model to map a string query to relevant docids for document retrieval.
They discover that tree-structured identifiers can establish structured information for candidate sets.
TIGER~\cite{rajput2023Tiger}, GPTRec~\cite{petrov2023GPTRec}, and P5~\cite{P5_zhangyongfeng, hua2023indexP5} employ text, user-item interactions, or historical sequences as prior knowledge, utilizing distinct indexing methods, e.g., RQ-VAE and SVD.
They consider the structure of identifiers in the construction phase, yet neglect it during the training process.
The user-item interactions are insufficient for the model to learn complex structured information.
\ourname optimizes structured information based on these findings.
We construct a balanced tree to map items to equal-length identifiers, ensuring semantic consistency at each layer and enabling more efficient inference with reduced tree depth.
We also introduce two contrastive learning tasks to the model training to aid in understanding the structure of the identifiers. 
Furthermore, \ourname achieves superior performance with just 1 transformer layer (with more parameters, \ourname can be better, as shown in Section~\ref{Appendix: parameter count}), whereas previous works required multiple layers, such as TIGER with 4 layers and P5 with 6 layers.


\subsection{Efficiency Analyses}
\label{sec: complexity}
We list complexity analysis of representative works with tree-structured identifiers in~\autoref{tab:complexity}, validating that structures of item identifiers in \ourname are more efficient.

Regarding space complexity, our emphasis is on the storage cost of identifiers. Given that current recommendation systems inherently require storing an item embedding table of size $Nd$ ($N$: the item count), our evaluation concentrates on the extra space introduced by identifiers.
In \ourname, identifiers' leaf tokens share embeddings of corresponding items; only non-leaf tokens add to additional space overhead.
Due to the structure of a balanced tree, the number of non-leaf tokens can be cumulatively calculated per layer: $1+k+k^2\cdots+\frac{\lceil\frac{N}{k}\rceil}{k}+\lceil\frac{N}{k}\rceil=\frac{k\lceil\frac{N}{k}\rceil-1}{k-1}$.
If $N$ is the power of $k$, then $\frac{k\lceil\frac{N}{k}\rceil-1}{k-1} = \frac{N-1}{k-1}$.
In our experiments, $k$ is set to $8$ or $16$.
Consequently, the additional space cost $\frac{N-1}{k-1} d$ introduced by identifiers is significantly smaller compared to $Nd$ (size of item embedding table).

To reduce time complexity, we leverage beam search during decoding.
In real-world applications, intermediary encoder outputs in \ourname can be efficiently precomputed and stored, as shown in previous works.
The bottleneck during inference is the beam search over identifiers.
Compared to TDM’s binary tree and RecForest's multiple trees, \ourname evidently shows an advantage in inference speed, as denoted in~\autoref{tab:complexity}.
Although DSI and NCI share a similar tree construction method with \ourname, their inference steps often amount to several times greater than \ourname. 
Due to their utilization of an imbalanced tree structure for identifiers, the max length of identifiers often is a constant multiple of $\log_k N$, and the max length critically influences inference speed.
Therefore, \ourname demonstrates a superior inference time relative to other tree-based and generative models.


\section{Experiment}

\begin{table}[t!]
\small
 \caption {Statistics of three public and one industrial datasets.}\label{tab: dataset} 
 \vspace{-5px}
{
\tabcolsep=0.17cm 
\begin{tabular}{lcccc}
\toprule
 Dataset & \#Users & \#Items &  \#Interactions & Density   \\ \hline
 Yelp & 31,668  & 38,048 & 1,561,406 & 0.130\%  \\
 News & 50,000  & 39,865 & 1,162,402  & 0.050\% \\ 
 Books & 459,133  & 313,966 & 8,898,041  & 0.004\% \\ 
 Micro-Video & 0.75 million  & 6.1 million & 85 million & 0.002\%  \\
 \bottomrule
\end{tabular}
}
\vspace{-10px}
\end{table}

\begin{table*}[h!]
    \small
	\caption{Performance comparison on three public datasets and an industrial dataset. The best and the second-best performances are denoted in bold and underlined fonts, respectively. * indicates significant improvements with \textit{p}-value < $0.05$.
    In this table, \ourname uses a single layer of encoder-decoder. The performance of SEATER is further improved with more layers of encoder-decoder, as shown in Section~\ref{Appendix: parameter count}.
 }
    \vspace{-5px}
	\label{table: main exp}
	\setlength{\tabcolsep}{1.5 mm}{
	\begin{tabular}{llccccccccccccc}
	\toprule
	\multicolumn{1}{l }{\multirow{2}{*}{Datasets}} & \multicolumn{1}{l }{\multirow{2}{*}{Metric}} & \multicolumn{7}{c}{\emph{Dual-encoder}}                 & \multicolumn{2}{c}{\emph{Tree-based Indexing}} 	& \multicolumn{3}{c}{\emph{Generative}} 					  \\ \cmidrule(l){3-9} \cmidrule(l){10-11} \cmidrule(l){12-14} 
	\multicolumn{1}{c}{}                          & \multicolumn{1}{c}{}                        & Y-DNN & GRU4Rec & MIND & ComiRec & SASREC & BERT4REC & Re4           & TDM & RecForest                                                     &GPTRec &TIGER &\ourname \\ \midrule
\multirow{6} * {Yelp}
	&NDCG@20  &0.0412 &0.0426 &0.0414 &0.0381 &0.0466 &0.0458 &0.0362 &0.0414 &0.0434   &0.0440 & \underline{0.0539}  &\textbf{0.0572}$^*$ \\
	&NDCG@50 &0.0613 &0.0628 &0.0611 &0.0581 &0.0699 &0.0688 &0.0595 &0.0610 &0.0646   &0.0653 & \underline{0.0769} &\textbf{0.0810}$^*$  \\
	&HR@20  &0.3366 &0.3467 &0.3502  &0.3263 &0.3711 &0.3746  &0.3263  &0.3493 &0.3503  &0.3487 & \underline{0.4087} &\textbf{0.4201}  \\
	&HR@50  &0.5396 &0.5507 &0.5409 &0.5319 &0.5799 &0.5774 &0.5468 &0.5439 &0.5512   &0.5527 &  \underline{0.5922} &\textbf{0.6118}$^*$ \\
	&R@20   &0.0511 &0.0529 &0.0522 &0.0508 &0.0594 &0.0590 &0.0462 &0.0524 &0.0549   &0.0559 &  \underline{0.0679} &\textbf{0.0720}$^*$ \\
	&R@50   &0.1045 &0.1071 &0.1046 &0.1034 &0.1206 &0.1204 &0.0976 &0.1040 &0.1110   &0.1121 &  \underline{0.1271} &\textbf{0.1353}$^*$ \\
\hline
\multirow{6} * {News}
	&NDCG@20  &0.0782 &0.0836 &0.0803 &0.0753 &0.0871 &0.0829 &0.0821 &0.0830 &0.0811  &0.0813 & \underline{0.0919}   &\textbf{0.0942}  \\
	&NDCG@50 &0.1047 &0.1114 &0.1076 &0.1011 &0.1142 &0.1072 &0.1107 &0.1067 &0.1068  &0.1065 &  \underline{0.1182} &\textbf{0.1225}$^*$ \\
	&HR@20  &0.3772 &0.3872 &0.3854 &0.3738 &0.3905 &0.3640 &0.3896 &0.3821 &0.3687  &0.3731 & \underline{0.4019}  &\textbf{0.4070} \\
	&HR@50  &0.5374 &0.5480 &0.5328 &0.5279 &\underline{0.5548} &0.5210 &0.5392 &0.5248 &0.5299  &0.5305 &  0.5531 &\textbf{0.5747}$^*$ \\
	&R@20   &0.1192 &0.1335 &0.1282 &0.1287 &0.1383 &0.1275 &0.1292 &0.1280 &0.1270  &0.1324 &  \underline{0.1408} &\textbf{0.1456}$^*$ \\
	&R@50   &0.2057 &0.2287 &0.2136 &0.2163 &\underline{0.2304} &0.2099 &0.2236 &0.2080 &0.2142  &0.2182 & 0.2292 &\textbf{0.2429}$^*$ \\
\hline
\multirow{6} * {Books}
	&NDCG@20  &0.0243 &0.0192 &0.0233 &0.0250 &0.0402 &0.0352 &0.0397 &0.0235 &0.0411   &0.0271 &  \underline{0.0468}  &\textbf{0.0592}$^*$ \\
	&NDCG@50 &0.0319 &0.0260 &0.0291 &0.0331 &0.0531 &0.0457 &0.0494 &0.0330 &0.0494   &0.0373 &  \underline{0.0573} &\textbf{0.0713}$^*$ \\
	&HR@20  &0.0977 &0.0820 &0.0861 &0.1169 &\underline{0.1661} &0.1374 &0.1455 &0.1101 &0.1347   &0.1181 &  0.1637 &\textbf{0.2006}$^*$ \\
	&HR@50  &0.1574 &0.1354 &0.1301 &0.1788 &\underline{0.2553} &0.2124 &0.2163 &0.1832 &0.1978   &0.1962 &  0.2380 &\textbf{0.2813}$^*$ \\
	&R@20   &0.0447 &0.0361 &0.0402 &0.0574 &\underline{0.0793} &0.0679 &0.0712 &0.0475 &0.0625   &0.0533 &  0.0766 &\textbf{0.0972}$^*$ \\
	&R@50   &0.0750 &0.0626 &0.0618 &0.0890 &\underline{0.1298} &0.1088 &0.1092 &0.0849 &0.0951   &0.0938 &  0.1179 &\textbf{0.1448}$^*$ \\
\hline
\multirow{6} * {Micro-Video}
	&NDCG@20  &0.0149 &0.0202 &0.0195 &0.0211 &0.0205 &0.0197 &\underline{0.0235} &0.0201 &0.0189   &0.0187 & 0.0230  &\textbf{0.0350}$^*$ \\
	&NDCG@50 &0.0186 &0.0254 &0.0244 &0.0289 &0.0253 &0.0238 &\underline{0.0293} &0.0240 &0.0214   &0.0221 &  0.0279 &\textbf{0.0406}$^*$ \\
	&HR@20  &0.1589 &0.1991 &0.1876 &0.2198 &0.2151 &0.1951 &\underline{0.2251} &0.1980 &0.1789   &0.1877 &  0.2223 &\textbf{0.2824}$^*$  \\
	&HR@50  &0.2728 &0.3287 &0.3027 &0.3567  &0.3424  &0.3159 &\underline{0.3624} &0.3077 &0.2898   &0.2928 &  0.3576 &\textbf{0.4037}$^*$ \\
	&R@20   &0.0118 &0.0186 &0.0175 &0.0199 &0.0191 &0.0167 &\underline{0.0231} &0.0190 &0.0178   &0.0166 &  0.0211 &\textbf{0.0310}$^*$ \\
	&R@50   &0.0269 &0.0383 &0.0357 &0.0403 &0.0391 &0.0338 &\underline{0.0451} &0.0342 &0.0322   &0.0331 &  0.0418 &\textbf{0.0566}$^*$ \\
\bottomrule
	\end{tabular}
	}
 \vspace{-5px}
\end{table*}

\subsection{Experimental Setup}

 We adhere to standard practices~\cite{TDM, comirec, NIPS22RecForest, Re4} for item retrieval by choosing suitable datasets, baselines, and evaluation metrics.

\subsubsection{Dataset.}

\autoref{tab: dataset} reports basic statistics of all the datasets.
We have selected the following three public datasets:
1) \textbf{Yelp}\footnote{\url{https://www.yelp.com/dataset}}: This dataset is adopted from the 2018 edition of the Yelp challenge. 
     The dataset encompasses business activities that occurred on the Yelp platform. 
2) \textbf{Books}\footnote{\url{http://jmcauley.ucsd.edu/data/amazon/}}: The Amazon review dataset~\cite{amazon_dataset} is one of the most widely used recommendation benchmarks. We adopt the `Books' subset. 
3) \textbf{News}\footnote{\url{https://msnews.github.io/}}: The MIND dataset is a benchmark for news recommendation. It is collected from the behavior logs of the Microsoft News website. We adopt the `MIND-small' subset. 

To evaluate our model in a real-world situation, we collected an industrial large-scale dataset from a commercial app.
4) \textbf{Micro-Video}: we randomly selected 0.75 million users who used a micro-video app over two weeks in 2023. The historical behaviors have been recorded. Unlike other public datasets, this industrial dataset has not undergone any filtering and exhibits high sparsity that aligns with real industrial scenarios.





\subsubsection{Evaluation Metrics \& Protocal}
\label{sec: exp metrics}

Following the common practices~\cite{comirec, NIPS22RecForest, Re4}, we divide each dataset into three parts, \ie training/validation/test sets by partitioning the users in a ratio of 8:1:1.
For evaluation, we take the first 80\% historical behaviors as context and the remaining 20\% as ground truth.
We strictly adhere to the evaluation framework in~\cite{comirec}. Please refer to~\cite{comirec} for details.
As for metrics, we employ three widely used metrics, including \textit{Hit Ratio} (HR), \textit{Normalized Discounted Cumulative Gain} (NDCG)\footnote{We compute the values based on the official definition of NDCG~\cite{NDCG_defination}, while a few existing works do not.}, and \textit{Recall} (R).
Metrics are calculated based on the top 20/50 recommended candidates (\eg HR@$20$).
We calculated them according to the ranking of items and reported the average results.

 Several baseline works calculate the NDCG scores \emph{incorrectly}, which has given rise to numerous discussions in their official GitHub Repositories, including Comirec (\url{https://github.com/THUDM/ComiRec/issues/6}), Re4 (line 179 in \url{https://github.com/DeerSheep0314/Re4-Learning-to-Re-contrast-Re-attend-Re-construct-for-Multi-interest-Recommendation/blob/main/src/model.py}) and RecForest (\url{https://github.com/wuchao-li/RecForest/issues/2}). 
 We fixed the issue by calculating IDCG (Ideal Discounted Cumulative Gain) using all positive items, instead of only considering the recalled positives used in those baselines which resulted in falsely higher NDCG in the above-mentioned papers.  

\subsubsection{Baseline and Implementation Details.}

We compare our model with SOTA models for item retrieval.
The mainstream models commonly adopt a \textit{\textbf{dual-encoder}} architecture:
(1) \textbf{YoutubeDNN} (abbreviated as Y-DNN)~\cite{youtubeDNN};
(2) \textbf{GRU4Rec}~\cite{GRU4REC};
(3) \textbf{MIND}~\cite{MIND_model};
(4) \textbf{ComiRec}~\cite{comirec} (the ComiRec-SA variant);
(5) \textbf{SASREC}~\cite{SASREC};
(6) \textbf{BERT4REC}~\cite{BERT4REC}; 
(7) \textbf{Re4}~\cite{Re4};
We also include models with \textit{\textbf{tree-based indexing }}: 
(8) \textbf{TDM}~\cite{TDM};
(9) \textbf{RecForest}~\cite{NIPS22RecForest};
Furthermore, we include latest \textit{\textbf{generative}} recommendation models:
(10) \textbf{GPTRec}~\cite{petrov2023GPTRec} (the GPTRec-TopK variant);
(11) \textbf{TIGER}~\cite{rajput2023Tiger} (implemented with 4 layers, the same as the original paper).

We also compare SEATER with \textbf{DSI} and \textbf{NCI}. Since DSI and NCI are primarily designed for search, leveraging textual query information as encoder input and built upon T5, they are not directly suitable for recommendation settings. 
SEATER's distinction with them stems from its approach to employing identifiers and the additional losses. 
To compare with DSI and NCI, we adapted SEATER by omitting the supplementary losses and adopting identifier structures from DSI and NCI. 
Detailed experimental results can be found in~\autoref{tab:ablation} and Section~\ref{sec:ablation}.

For baselines, we tune the hyper-parameters following the suggestions in the original papers.
For SASREC and the five dual-encoder models, we train them using the sampled softmax loss~\cite{comirec}, commonly adopted for the matching phase, setting the negative sample size to 1280.
For other baseline models, we adopt the loss functions and training procedures described in the original papers.
For all models, the dimension of item embeddings is set to 64.
All the dual-encoder and transformer-based models make predictions based on \textit{\textbf{brute-force}} retrieval, which involves calculating the probability over all items.
For fair competition, we use the same item embeddings to build indexes for both \ourname and RecForest.
Considering that TIGER requires using item text information to construct codebooks, and only the MIND dataset provides texts of items, we leveraged the SASREC embeddings for other datasets.
As for TDM, RecForest, and~\ourname, they predict based on \textit{\textbf{beam search}} over item indexes, where we set the beam size to $50$ for all of them.

We tune the hyper-parameters of \ourname as follows: the number of layers for encoder and decoder is set to 1; the values of loss coefficients, \ie $\lambda_a$ and $\lambda_r$, are searched from [1e-2, 9e-2] with step 2e-2; the $L_2$ regularization weight is searched from [1e-4, 1e-5, 1e-6, 1e-7];
the number of tree branches $k$ is searched in [$2,4,8,16,32$];
the number of sampled identifiers $q$ is set to 4;
the margin value $\beta$ is searched in [$0.01, 0.001, 0.0001$].
We use Adam~\cite{Adam} for optimization with a learning rate of $0.001$, and adopt the early stop training to avoid over-fitting.
We provided code and data at an this link (\textcolor{magenta}{\url{https://github.com/Ethan00Si/SEATER_Generative_Retrieval}}).


\subsection{Overall Performance}

\begin{table*}[ht!]
	\small
	\centering
	\caption{Ablation study on four datasets.
        We assess the proposed two losses and the designed identifiers.
        Each loss contributes positively to the model, as shown in the middle four rows.
        Using DSI and NCI's decoder shows worse performance compared to SEATER's decoder structure, as shown in the last two rows.
        }
	\vspace{-5px}
	\label{tab:ablation}
	\scalebox{1}{
	\setlength{\tabcolsep}{1mm}{
	\begin{tabular}{lcccccccccccc}
	\toprule
	\multicolumn{1}{c}{\multirow{2}{*}{\textbf{Variants}}} & \multicolumn{3}{c}{Yelp}                 & \multicolumn{3}{c}{News}  	 & \multicolumn{3}{c}{Books}	 & \multicolumn{3}{c}{Micro-Video}					         \\ \cmidrule(l){2-4} \cmidrule(l){5-7} \cmidrule(l){8-10} \cmidrule(l){11-13} 
	\multicolumn{1}{c}{}                                   & NDCG@50& HR@50 & R@50 & NDCG@50& HR@50 & R@50  & NDCG@50& HR@50 & R@50 & NDCG@50& HR@50 & R@50 \\ \midrule
	(0) \ourname                                           & \textbf{0.0810}  & \textbf{0.6118}  & \textbf{0.1353} & \textbf{0.1225}  & \textbf{0.5747}  & \textbf{0.2429} & 0.0713  & \textbf{0.2813}  & \textbf{0.1448} &\textbf{0.0406} &\textbf{0.4037} &\textbf{0.0566} \\ \cmidrule(r){1-1}
	(1) w/o $\mcal{L}_{\mrm{rank}}$ \& $\mcal{L}_{\mrm{ali}}$        & 0.0736     & 0.5920      & 0.1241  &0.1142 &0.5546 &0.2309 &\underline{0.0715} &0.2770 &0.1411  &0.0390 &0.3856 &0.0514   \\
	(2) w/o $\mcal{L}_{\mrm{rank}}$            & 0.0748       & 0.6029                  & 0.1266  &0.1201 &0.5687 &0.2386 &0.0710 &0.2802 &0.1433  &0.0394 &0.3947 &0.0531       \\
	(3) w/o $\mcal{L}_{\mrm{ali}}$       & \underline{0.0782}       & 0.6054                 & \underline{0.1317}     &0.1167 &0.5548 &0.2335 &\textbf{0.0721} &0.2779 &0.1427  &0.0395 &0.3931 &0.0534    \\
	(0) + w/o $\mcal{L}_{\mrm{rank}}$ for negatives        & 0.0760        & \underline{0.6063}           & 0.1298   &\underline{0.1208} &\underline{0.5717} &\underline{0.2401} &  0.0714 &\underline{0.2807} &\underline{0.1441} &\underline{0.0402} &\underline{0.3965} &\underline{0.0544}  \\\cmidrule(r){1-1}
	(1) + DSI Identifiers       & 0.0551      & 0.5019                & 0.0952 & 0.0947 & 0.4862 & 0.1919 &0.0408 &0.1908 &0.0902  &0.0225 &0.2727 &0.0313       \\ 
	(1) + NCI Identifiers        & 0.0618        & 0.5316             & 0.1053 & 0.1047 & 0.5145 & 0.2067 &0.0565 &0.2128 &0.1024  &0.0343 &0.3074 &0.0365   \\  \bottomrule
	\end{tabular}
	}}
	\vspace{-5px}
\end{table*}

\autoref{table: main exp} reports the overall performance on the four datasets. 
We have the following observations:

\noindent$\bullet$ \textbf{\ourname achieves the best performance on all datasets.}
\ourname consistently outperforms baselines of various types by a large margin.
Specifically, the relative improvements in R@50 on the Yelp, News, Books, and Micro-Video datasets are $6.45\%$, $5.43\%$, $11.56\%$, and $25.50\%$, respectively.
These results underscore \ourname's effectiveness.

\noindent$\bullet$ \textbf{\ourname significantly outperforms dual-encoder models and tree-based indexing models.}
Compared with dual-encoder models like SASREC, \ourname's improvement primarily stems from its generative decoding method, which models interaction probabilities more precisely than the inner product used by dual-encoder models. 
\ourname's improvement over models like ComiRec and Re4, which use multiple vectors to express user interests, confirms that expressing user interests through behavioral sequences provides a more comprehensive and thorough representation than using compressed vectors.
Additionally, \ourname surpasses models employing contrastive learning to enhance user interest representation, such as Re4, validating the effectiveness of optimizing identifiers as indices.


\noindent$\bullet$ \textbf{\ourname surpasses other generative methods in overall comparisons.}
TIGER utilizes 4 layers of encoder-decoder, while \ourname, with only 1 layer in this table, still achieves superior performance after significantly reducing resource consumption, validating the efficiency of \ourname.
Moreover, on sparser and larger datasets, the improvement of \ourname is larger. The results indicate \ourname is better suitable for industrial applications. 
The improvement over TIGER and GPTRec also validates the effectiveness of enhancing the structure of item identifiers in \ourname, i.e., the balanced tree structure and contrastive learning tasks for understanding structural information.

\subsection{Ablation Study}  
\label{sec:ablation}

We evaluated the performance impact of \ourname's components via an ablation study. 
The results are reported in~\autoref{tab:ablation}.

To assess the efficacy of the proposed losses $\mcal{L}_{\mrm{rank}}$ and $\mcal{L}_{\mrm{ali}}$, we test the following variants: Variant (1) excludes both loss terms; while Variants (2) and (3) remove $\mcal{L}_{\mrm{rank}}$ and $\mcal{L}_{\mrm{ali}}$ respectively, to study their contributions.
Both loss functions demonstrate a favorable influence on the model performance. 
Removing either one individually leads to a decline in overall performance.
Furthermore, we advanced our investigation by eliminating the ranking among negative samples within the ranking loss $\mcal{L}_{\mrm{rank}}$, as shown in~\autoref{eq: ranking loss}.
This led to the creation of Variant (0) + w/o $\mcal{L}_{\mrm{rank}}$ for negatives.
The performance of this Variant exceeds that of Variant (2) but falls short of Variant (0). This observation suggests that the inclusion of ranking among negative samples enhances the model's capability.
These phenomenons illustrate that these two loss functions aid the model in comprehending the tree structure of identifiers, such as the inter-token relationships and hierarchies within the tokens.

To compare \ourname with DSI and NCI, we created two variants (1) + DSI Identifiers and (1) + NCI Identifiers, based on Variant (1). These variants leverage the imbalanced tree construction, token embedding allocation methods, and decoder structures from DSI and NCI. In specific, (1) + DSI Identifiers assigns $k$ unique token vectors for a $k$-ary imbalanced tree. (1) + NCI Identifiers employs a layer-wise assignment of token embeddings which allocates $kL$ unique token vectors for a $k$-ary imbalanced tree of depth $L$. (1) + NCI Identifiers also employs the PAWA decoder following the NCI paper. Apart from these, all other variables, such as the embedding used in building identifiers and the number of model layers, remain consistent with \ourname for a fair comparison.  A significant decline in performance is noted in both variants compared with Variant (1), exhibiting an average performance drop exceeding $10\%$. This phenomenon validates the effectiveness of a balanced structure and suggests that shared embeddings for identifier tokens limit performance, especially in large-scale recommendation scenarios.

\subsection{Study on Item Identifiers}

\subsubsection{Impact of Different Item Identifiers}
\label{sec: exp item identifiers}

To verify our statements in Section~\ref{sec: identifiers},  
we investigated the impact of tree balance and the utilization of different embeddings on model performance.
As for tree balance, we utilized constrained k-means for a balanced tree and k-means for an imbalanced tree.
As for embeddings used for hierarchical clustering, we explored employing SASREC's item embeddings, obtaining embeddings from items' textual descriptions using BERT, and randomly initialized embeddings.
Owing to the exclusive presence of items' textual descriptions in the News dataset, we conducted experiments on this particular dataset.
For BERT embeddings, we concatenated the category, subcategory, title, and abstract of the news articles within the News dataset to form the input for BERT. 
Subsequently, we extracted the embedding of the [CLS] token and employed it as the corresponding item embedding.
For randomly initialized embeddings, we create them with random samples from a uniform distribution.

\begin{figure}[t]
    \centering
    \includegraphics[width=0.75\linewidth]{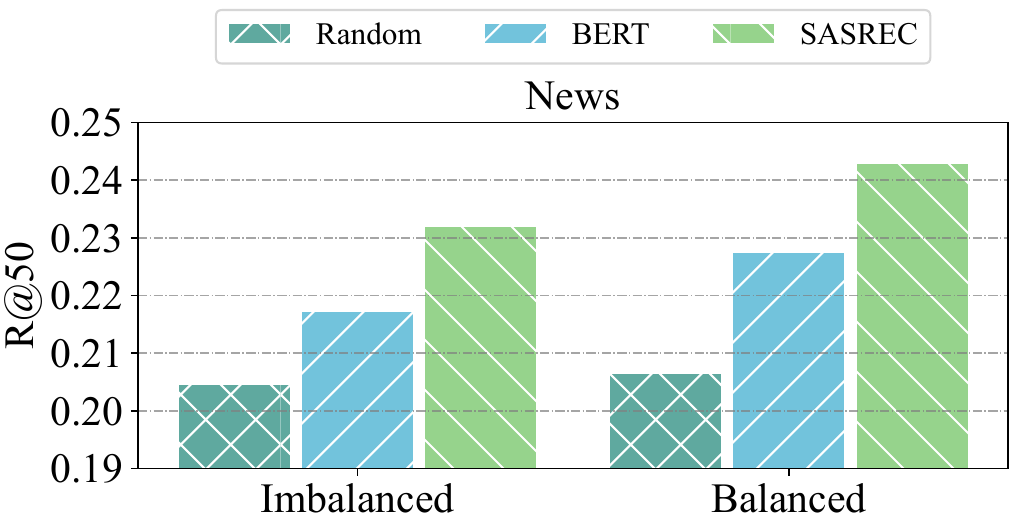}
    \vspace{-10px}
    \caption{Different methods to construct identifiers.
    The collaborative filtering information and balanced structure make identifiers more informative.
    }
    \label{fig: identifiers construction}
    \vspace{-10px}
\end{figure}

The results are shown in~\autoref{fig: identifiers construction}.
We observed that utilizing a balanced tree yields performance improvements compared to using an imbalanced tree when employing identical sources of item embeddings.
This is because a balanced tree ensures that tokens at the same level carry consistent semantic granularity, thereby capturing similar semantics within one layer.
We also observed that employing the item embeddings from SASREC yields optimal performance. This is attributed to its alignment with collaborative filtering information, rendering it more suitable for recommendation tasks.
Utilizing random identifiers yields the poorest performance, as it fails to impart any information gain to the identifiers.

\subsubsection{Effect of Branch Number $k$}
\label{sec: exp of branch number k}


\begin{figure}
    \centering
    \begin{subfigure}{0.45\linewidth}
        \includegraphics[width=\textwidth]{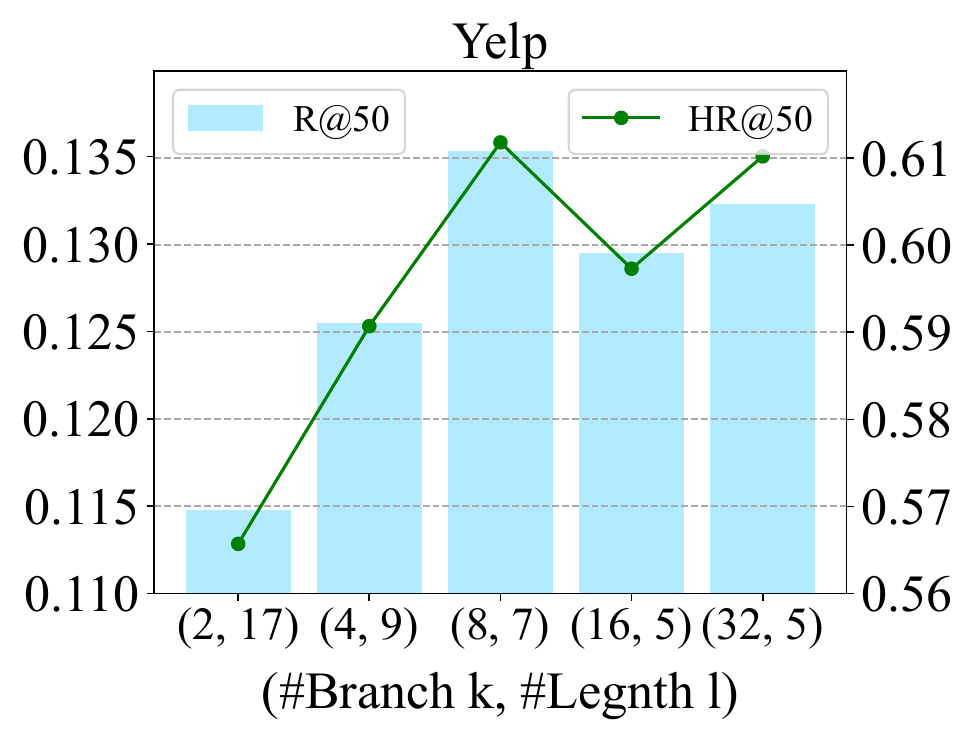}
    \end{subfigure}
    \begin{subfigure}{0.45\linewidth}
        \includegraphics[width=\textwidth]{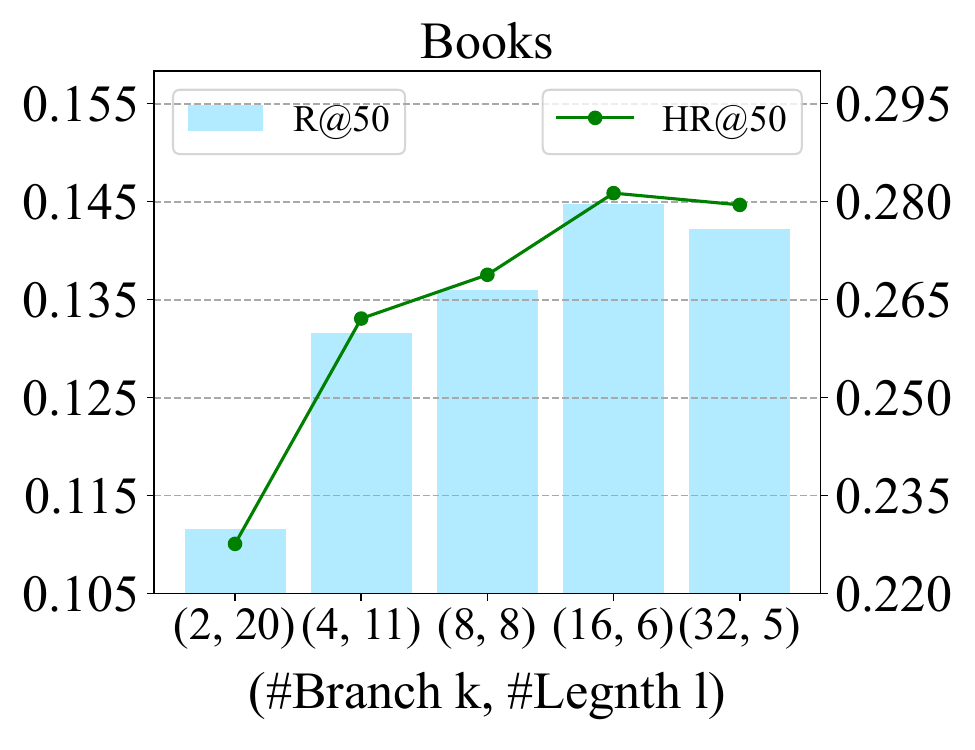}
    \end{subfigure}
    \vspace{-10px}
    \caption{Impact of branch number $k$, ranging from $2$ to $32$, in terms of R@50 and HR@50.
    The corresponding identifier length $l$ is also annotated. 
    }
\label{fig: branch}
\vspace{-10px}
\end{figure}

The variation in branch number $k$ leads to a corresponding alteration in the length $l$ of the item identifier. As $k$ increases, $l$ decreases.
We adjusted the size of $k$ and recorded the corresponding values of $l$ along with the model's performance. 
This experiment employed two datasets, Yelp and Books, with varying item quantities.
The results are illustrated in~\autoref{fig: branch}.
We observe that as $k$ increases from 2 to 8 on the Yelp dataset (or 2 to 16 on the Books dataset), the model's performance reaches its peak, while further increasing $k$ leads to a decline in performance.
The performance improvement resulting from increasing $k$ can be attributed to the reduction in identifier length $l$. 
As the beam search for inference cannot guarantee the selection of the correct next tokens at every step, a greater number of beam search steps (larger $l$) increases the probability of ultimate errors (due to cumulative errors).
The decline in model performance as $k$ increases from 8 to 32 on the Yelp dataset (or 16 to 32 on the Books dataset) is attributed to the fact that $l$ remains relatively unchanged while $k$ continues to increase.
The beam search selects the top $b$ options from $b*k$ candidate results at each step. 
The increase of $k$ amplifies the difficulty of beam search at every step, while $l$ results in a relatively unchanged total number of steps.
Hence, both large and small values of $k$ can lead to a decline in model performance.

\subsection{Analysis on Parameter Count}
\label{Appendix: parameter count}
We investigated the impact of the number of encoder-decoder layers.
We observed distinct patterns across datasets of varying scales.
The experiments were conducted on a small-scale dense dataset, Yelp, and a large-scale sparse dataset, Books.

\begin{figure}[t]
    \centering
    \begin{subfigure}{0.45\linewidth}
        \includegraphics[width=\textwidth]{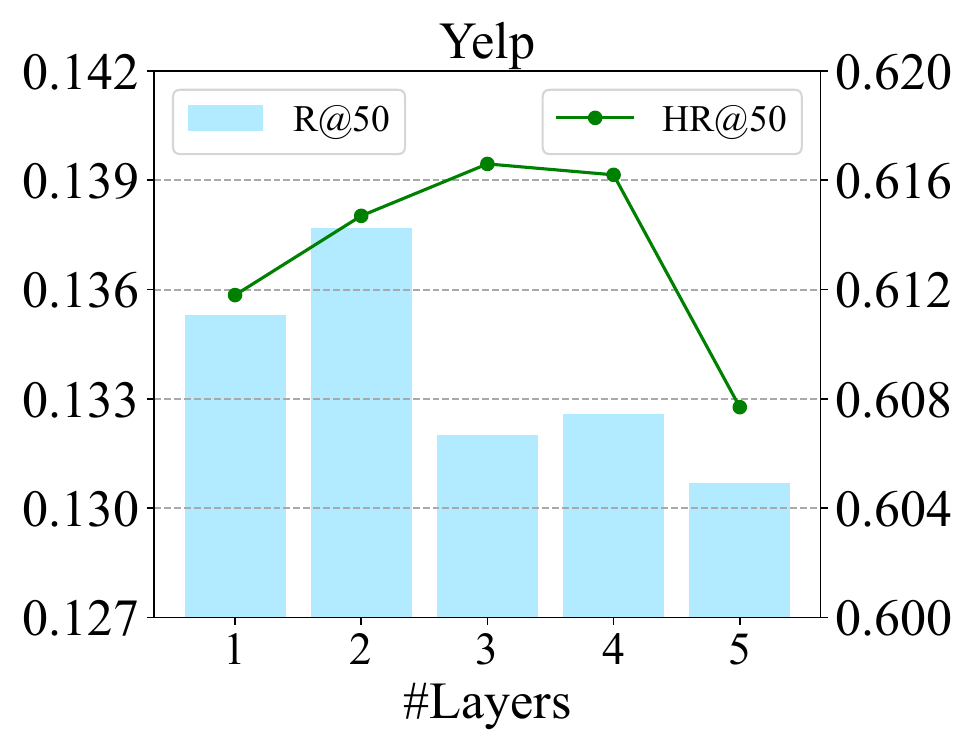}
    \end{subfigure}
    \begin{subfigure}{0.45\linewidth}
        \includegraphics[width=\textwidth]{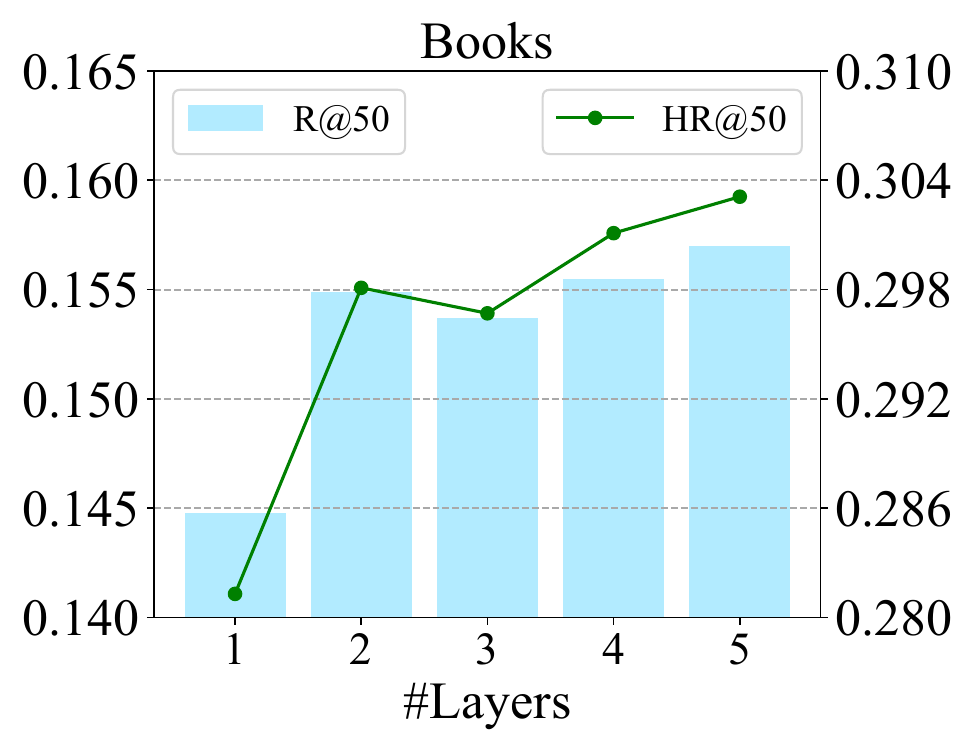}
    \end{subfigure}
    \vspace{-15px}
\caption{Analysis of the number of transformer layers.}
\label{fig: hyperpara n_layers}
\vspace{-17px}
\end{figure}

As shown in the left part of~\autoref{fig: hyperpara n_layers}, when the number of layers increases from 1 to 3, the performance is further improved on the Yelp dataset.
However, as the model's depth continues to increase, performance gradually deteriorates.
We discovered that this phenomenon is attributed to the smaller scale of the Yelp dataset, where overfitting occurs as the model's depth increases.
As shown in the right part of~\autoref{fig: hyperpara n_layers}, by increasing the model depth on the Books dataset, there is a continuous improvement in the model's performance. We posit that this is because a larger parameter count enhances the model's expressive capability on this dataset of a larger scale.
We leave deeper models, such as those with 12 layers, for future work.
Increasing the number of layers leads to a linear growth in computational complexity. This implies that the computational resources consumed by 2-layer and 3-layer models can be roughly considered as 2 times and 3 times that of a 1-layer model, respectively.
Thus, considering the lower speed of deeper models, we find that 1-layer models can strike a satisfactory balance between performance and efficiency, as they have already attained peak performance compared with other baselines.
Deepening the number of model layers is a promising direction for future research.

\section{Conclusion}
In this paper, we propose a generative retrieval model, namely \ourname, for recommendation. With contrastive learning tasks and balanced identifiers, \ourname achieves both efficiency and effectiveness by enhancing the structure of item identifiers.
With the help of two contrastive learning tasks, \ourname captures the nuances of identifier tokens, including unique semantics, hierarchies, and inter-token relationships.
Specifically, \ourname aligns token embeddings based on their hierarchical positions using the infoNCE loss and directs the model to rank similar identifiers in desired orders using the triplet loss.
\ourname exploits a balanced $k$-ary tree structure for identifiers, leading to rational semantic space allocation and fast inference speed.
This balanced structure maintains semantic consistency within the same level while different levels correlate to varying semantic granularities.
Detailed analyses of time and space complexities validate the efficiency of the proposed model, enabling its application on large-scale retrieval.
Extensive experiments on three public datasets and an industrial dataset verify that \ourname consistently outperforms SOTA models of various types.


\bibliographystyle{ACM-Reference-Format}
\bibliography{main}



\end{document}